# Towards a perfect fixed point action for SU(3) gauge theory

Thomas A. DeGrand[a], Anna Hasenfratz[a] *, Peter Hasenfratz[b], Ferenc Niedermayer [b] and Uwe Wiese[c]

[a]Physics Department, University of Colorado, Boulder, CO 80309

[b]Institute for Theoretical Physics, University of Bern, Sidlerstr. 5, CH-3012 Bern

[c]Physics Department, M. I. T., Cambridge, MA 02139

We present an overview of the construction and testing of actions for SU(3) gauge theory which are approximate fixed points of renormalization group equations (at $\beta \to \infty$). Such actions are candidates for use in numerical simulations on coarse lattices.

## 1. INTRODUCTION

"Perfect actions" are actions which are defined on a lattice but show no lattice artefacts. Any action on a renormalized trajectory of any renormalization group transformation is perfect, free of lattice artifacts. If one can find such an action that, in addition to being perfect, is reasonably local as well, than the computational overhead associated with simulating the more complex action can be easily compensated by its improved scaling behavior. This program was successfully carried out for the 2 dimensional O(3) $\sigma$ model[1].

We have been trying to construct and test perfect actions for SU(3) pure gauge theory. Our ultimate goal is to find a perfect action for QCD with fermions where scaling violations are more serious. This talk summarizes some of our results to date.

Lattice regularization introduces various artefacts like scaling violation or rotational symmetry breaking. The distortion caused by lattice artefacts can depend strongly on the specific choice of lattice action. Most numerical calculations to date use the Wilson plaquette action. This action is the simplest and computationally the fastest per site but the continuum limit is probably reached only at fairly small lattice spacing and, consequently, typical pure gauge or quenched calculations are done on spatial volumes $32^3$ or larger.

The alternative to decreasing the lattice spacing in controlling lattice artefacts is to use an improved action. Although the history of improved actions is long they have had a limited effect on actual calculations until now. The reason is that the computational overhead of simulating these actions was not compensated by the possible improvements obtained in the scaling properties. A new action obtained by combining a perturbatively improved action with non-perturbative corrections has been proposed at this conference[2]. The theoretical background of our method is completely different from that of [2].

Our goal here is to find a systematic way to construct and test a perfect action which is largely free of lattice artifacts. In the following we first discuss scaling tests for SU(3) gauge theory. Next we describe how a fixed point action can be obtained and investigate a candidate action.

## 2. SCALING TESTS

Scaling means that all physical dimensional quantities show the same functional dependence on the gauge coupling. It is different from asymptotic scaling where in addition to scaling we require that this functional dependence be described by the 2-loop $\beta$ function.

Scaling violations can be different for different quantities. The ratio of certain observables might show scaling at smaller correlation length than others. Scaling requires the universal behavior of *all* observables.

---
*Talk delivered by Anna Hasenfratz

When we started this project we thought that data to show scaling or the lack of it for the Wilson action would be readily available in the literature. Unfortunately that is not the case. Scaling tests can be very sensitive to finite volume effects and the systematic errors associated with infinite volume extrapolations make many of the existing data unusable. This is especially so for $\beta > 6.0$.

We have decided calculate physical observables for our scaling test in fixed, finite physical volumes. This way we can avoid the infinite volume extrapolation. For one scaling test we consider the quantity $G = \sqrt{\sigma(L)}/L$ where $\sigma(L)$ is the string tension on an $L^3$ volume computed from the exponential fall-off of the Polyakov line correlator (torelon mass). If evaluated in *constant physical volumes*, G is independent of the bare coupling in the scaling (continuum) limit. Any variation of G is due to lattice artefacts. Constant physical volumes can be achieved, for example, by evaluating G at the critical coupling of the finite temperature phase transition $aT_c = 1/N_T$ in volumes $L^3 = (rN_T)^3$ where $r$ is some conveniently chosen aspect ratio. Figure 1 shows $G(L)$ with $r = 2.0$ as the function of $aT_c$ for the Wilson action. The plot contains data for $N_T$ between 2 and 14 spanning the coupling constant range from $\beta = 5.1$ to $\beta = 6.4$. The torelon masses for $\beta \geq 5.9$ ($N_T \geq 6$) are from published data[3]. The large errors are due to the uncertainty of the infinite volume finite temperature phase transition. The Wilson action appears to show an approximately 10% scaling violation between $N_T = 2$ and 8.

To avoid extrapolating $\beta_c$ to infinite volume one can set the scale by the finite temperature phase transition at finite, constant physical volumes $N_S^3 N_T$, $N_S = rN_T$. Strictly speaking there is no phase transition in finite volume and we have to define what we consider $\beta_c(N_S)$. We found that the Columbia group's definition [4] of $\beta_c$ is the most reliable on small volumes. It is based on monitoring the phase of the Polyakov loop and requiring 50-50% occupation time for both the deconfined and confined vacua.

This quantity shows smaller scaling violations than the previous one. It is displayed as octagons in Fig. 2. Surprisingly (to us) the Wilson action data for this observable is consistent with scaling for $N_T \geq 2$.

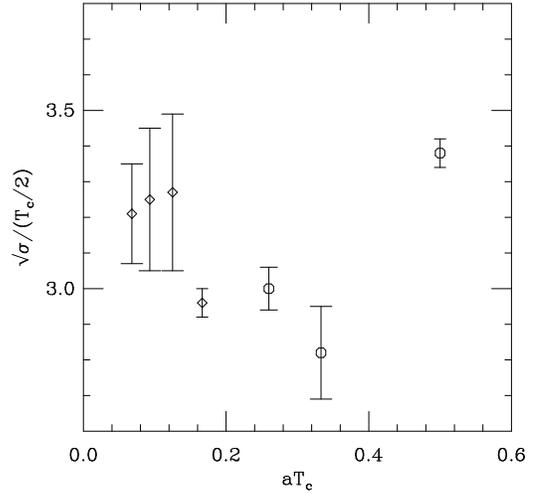

Figure 1. Scaling test for the Wilson action; $T_c$ is defined in infinite spatial volumes.

Alternatively one could use the potential evaluated at fixed physical volumes and temperature to study scaling. However, the potential is derived from a correlator of operators and its behavior is sensitive to the properties of operators in a way that a mass measurement is not. When scaling holds the quark-antiquark potential is rotationally invariant even at small distances, but that might not be evident if one measures an operator that creates, instead of one localized fermion, some spatially extended quark distribution. The naive Polyakov loop is such an operator. The perfect potential operator is an operator that describes one heavy quark at a definite location. It is in principle a combination of Polyakov lines and Wilson loops and depends on the action itself. Fig. 3 shows $V(r)/T_c$ as the function of $rT_c$ for the Wilson action at two different values of the lattice spacing. Crosses and octagons correspond to measurements at $\beta_c(N_T = 2)$ and $\beta_c(N_T = 4)$ on $(2N_T)^3(1.5N_T)$ lattices respectively. The potentials, measured with naive Polyakov loops, are normalized so $V(r = 1/T_c) = T_c$. The lack of scaling is evident. We would like to emphasize that lack of rotational invariance of the direct Polyakov loop operators does not mean lack of scaling and, alternatively, perfect rotational invariance does not necessarily mean perfect scaling either. Scaling violation is evident from the overall difference of the two potentials.

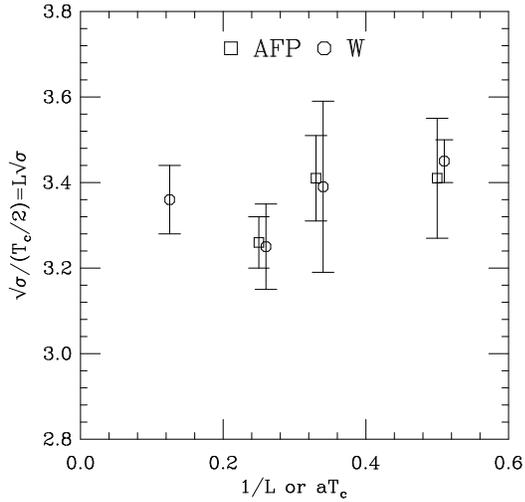

Figure 2. Scaling test for the Wilson action (octagons) and AFP action (squares); $T_c$ is defined in constant physical volumes.

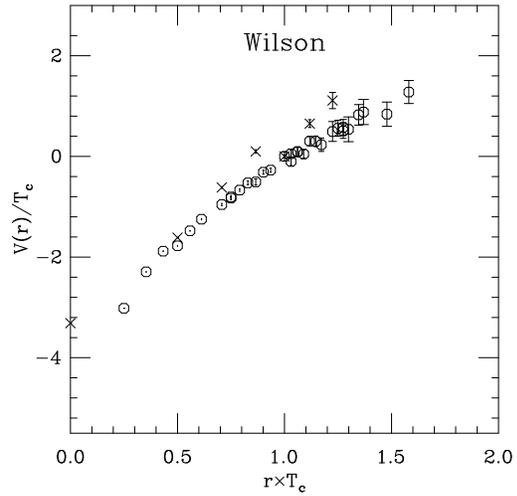

Figure 3. Potential $V(r)/T$ vs. $rT$ for the Wilson action at $\beta_c(N_T = 2)$ (crosses) and $\beta_c(N_T = 4)$ (octagons).

## 3. THE CLASSICAL FIXED POINT ACTION

### 3.1. Perfect actions

Consider an SU(3) gauge theory defined on a hypercubic lattice. The partition function is

$$Z = \int DU e^{-\beta S(U)}, \qquad (1)$$

where $\beta S(U)$ is some lattice representation of the continuum action. It is a combination of the products of the $U_\mu(n) = e^{-a g A_\mu(n)}$ SU(3) matrices along arbitrary closed loops in arbitrary representation.

Under repeated real space renormalization group transformations the action moves in a multidimensional parameter space. On the critical surface it runs into a fixed point (FP). The FP of the SU(3) gauge theory is at $\beta = \infty$ and has one marginally relevant and infinitely many irrelevant directions. The trajectory which leaves the FP along the marginal direction is the renormalized trajectory, RT.

The FP and any point on its RT describe perfect actions. There are no lattice artefacts along the RT. The FP/RT, while describing perfect actions, can be very complicated containing many non-negligible interaction terms. For practical purposes one needs a sufficiently short ranged FP/RT that can be parametrized by a few operators, at least at correlation lengths that are relevant to numerical simulations.

In the present work we concentrate on locating a FP on the critical surface. Since the RT leaves the critical surface perpendicular to it the FP describes the whole RT close to the critical surface. We assume that it stays close at smaller correlation lengths as well. This is an uncontrolled assumption and is potentially the most serious source of error for this project.

### 3.2. The fixed point action

The location of the FP is not unique. It depends on the particular choice of renormalization group transformation. We worked with two dif-ferent scale 2 block transformation with tunable free parameters. Both transformations can be defined as

$$e^{-\beta'S'(V)} = \int DU e^{-\beta(S(U)+\kappa T(U,V))} \quad (2)$$

where U is the original, V is the blocked link variable and $T(U,V)$ is the blocking kernel that defines the transformation.

The first transformation we used is a modified version of the Swendsen transformation where the central link is given a different weight than the staples. The other transformation uses an APE type smearing to create fuzzy links. The blocked link is chosen around the product of two fuzzy links with a Boltzman factor as indicated in Eqn. 2.

On the critical surface, in the limit $\beta \to \infty$ Eqn. 2 reduces to a saddle point problem giving

$$S'(V) = min_{\{U\}} \left( S(U) - \kappa T(U,V) \right). \quad (3)$$

The FP of the transformation is determined by the equation

$$S^*(V) = min_{\{U\}}(S^*(U) - \kappa T(U,V)). \quad (4)$$

For smooth configurations this equation can be solved perturbatively. At smaller correlation lengths the configurations are not smooth any more and the solution of the equation has to be obtained by numerical minimization. The tunable parameters can be used to find the most local fixed point within its class.

### 3.3. Parametrization of the fixed point action

We have used a parametrization of the action based on powers of the traces of the loop products $U(C) = \Pi_C U_\mu(n)$ where $C$ is an arbitrary closed path

$$\begin{aligned}S(U) = \frac{1}{3}\sum_C \quad & (c_1(C)(3 - ReTr(U(C)) \\ & + \; c_2(C)(3 - ReTr(U(C))^2 + ... \\ & + \; d_2(C)(ImTr(U(C))^2 + \; ...)\,. \end{aligned} \quad (5)$$

Since we are searching for a short range fixed point first we restricted the sum in Eqn. 5 to the 28 loops that are length 10 or less and fit into a $2^4$ hypercube. For practical purposes one needs to reduce the parameter space even further. For numerical simulations we consider the projection of the action to a subspace of two operators in 3 representations. We chose the plaquette and the twisted 6 link operator $(x,y,z,-x,-y,-z)$ for this restricted parametrization.

### 3.4. The fixed point at the quadratic and cubic level

For smooth configurations the RG equation can be solved analytically. At the quadratic level Eqn. 4 translates into a relation between 2-point correlation functions. By repeating the transformation the equation can be iterated to find the fixed point that is expressed in terms of link variables. In terms of the loop parametrization this will restrict the coefficients of the linear terms $c_1(C)$. As there are many more loops than links the translation of the quadratic fixed point into the loop parametrization is not unique. For example only 7 of the 28 loops in the $2^4$ hypercube are independent at the quadratic level.

It is possible to go one step further and solve the fixed point equation at the cubic level. The cubic level does not restrict any of the higher representation terms in the action but constrains the $c_1$ coefficients of 15 out of the 28 loops.

Since these calculations are analytic we can easily investigate the properties of different renormalization group transformations and find the one that gives the most local fixed point action. The best FP action is quite local especially for the second type of transformation and is dominated by the plaquette term in both cases.

### 3.5. Numerical minimization

To go beyond leading order we solved the FP equation numerically. First we generate an SU(3) configuration $\{V\}$. This is our coarse configuration. Next we choose an action $S_0$ that is sufficiently simple for minimization but close to the fixed point action. The minimization of Eqn. 3 gives the fine configuration $\{U_0\}$ that blocks into $\{V\}$

$$\begin{aligned}S'(V) = \; & min_U(S_o(U) + T(U,V)) \\ = \; & S_0(U_0) + T(U_0,V). \end{aligned} \quad (6)$$

If $S_0$ is close to the fixed point action $S^*$ than $U_0$ is close to $\{U^*\}$ and at leading order

$$S^*(V) = S^*(U_0) + T(U_0, V)$$
$$= S^*(U_0) + (S'(V) - S_0(U_0)). \quad (7)$$

This equation can be used to calculate the value of the fixed point action on the course configuration $\{V\}$ if $S^*$ is known on the fine configuration $\{U_0\}$. Using the parametrization for $S^*$ and repeating the above procedure on many $\{V\}$ configurations leads to a set of linear equations for the coefficients $c_i^*$.

On very fine configurations the FP action is well described by its quadratic form and can be used to evaluate $S^*(U_0)$ in Eqn. 7. This in turn will give $S^*(V)$ on coarser configurations that, if necessary, could be used to calculate $S^*$ in even coarser configurations, etc.

The procedure could fail if $S_0$ in the original minimization was not close to the FP action. We found that by carefully adjusting the couplings of $S_0$ one can make the difference $S^*(U_0) - S_0(U_0)$, that measures this correction, arbitrary small.

## 4. THE FEW PARAMETER FP ACTION

Using $O(50)$ parameters we could get a good parametrization of $S^*$ down to correlation length $\xi = O(1)$. However this action is too complicated for numerical simulation. While one cannot expect that a few parameter action will correctly describe the FP action over a wide range of correlation lengths, it is possible to get a good fit for the FP action with only a few parameters if we restrict ourselves to a smaller correlation length range. For that purpose we generated coarse configurations with the Wilson action between $\beta = 5.0$ and $6.0$ and fitted $S^*(V)$ with two operators in 3 representations as described in Sect.3.3. The coefficients of the approximate FP (AFP) action is given in table 1. It is not possible to find a FP action using only operators in the fundamental representation.

The AFP fits the minimization data well for configurations generated with Wilson $\beta \geq 5.7$. Below that the quality of the fit deteriorates indicating that other types of operators are needed to describe the FP action on configurations with smaller correlation length. We expect the second type of block transformation, which gives a more local quadratic fixed point, will have better properties at small correlation lengths.

### 4.1. Properties of the AFP action

The AFP describes the FP on the critical surface. The properties of the AFP action at finite $\beta$ small correlation length range cannot be predicted. A detailed scaling test is needed to show the properties of the AFP action.

We have written programs to simulate actions like the ones shown in table 1. We have a version for serial machines which uses a combination of overrelaxation and Metropolis updating, and a parallelized version using the hybrid Monte Carlo algorithm. The time to update a link in the action of Table 1 is about 9 times the update time of our Wilson code.

We have been performing one scaling test, a measurement of the torelon mass on lattices of constant aspect ratio. We set the scale for the scaling test by the finite temperature phase transition in finite physical volumes. While one does not expect improved asymptotic scaling for the AFP action it is nevertheless interesting to check the perturbative scaling properties and the value of the $\Lambda$ parameter of the AFP action. Figure 4 shows $T_c/\Lambda$ both for the Wilson and AFP actions. The AFP action not only shows asymptotic scaling from $N_T = 2$ but the value of its $\Lambda$ parameter is about a factor of ten larger than the Wilson one and therefore much closer in value to the continuum $\Lambda$ parameters.

As for the quantity $G = \sqrt{\sigma(L)}/T_c$, the AFP action appears to scale over the range $a > 1/(2T_c)$, as shown in Fig. 2 (squares).

Finally, we have measured the potential from the correlation function of Polyakov loops. In Fig. 5 we show $V(r)/T_c$. The notation is the same as for the Wilson action in Fig. 3. Compared to the Wilson action the AFP action shows much better rotational invariance and, what is more important, it shows better scaling between $N_T = 2$ and 4. As we mentioned earlier to see the restoration of rotational invariance one needs to use perfect operators in the potential measurement.

6

Table 1
Couplings of the AFP action currently under investigation.

| operator | $c_1$ | $c_2$ | $c_3$ | $c_4$ |
| --- | --- | --- | --- | --- |
| $c_{plaq}$ | .523 | .0021 | .0053 | .016 |
| $c_{6-link}$ | .0597 | .0054 | .005 | -.0006 |

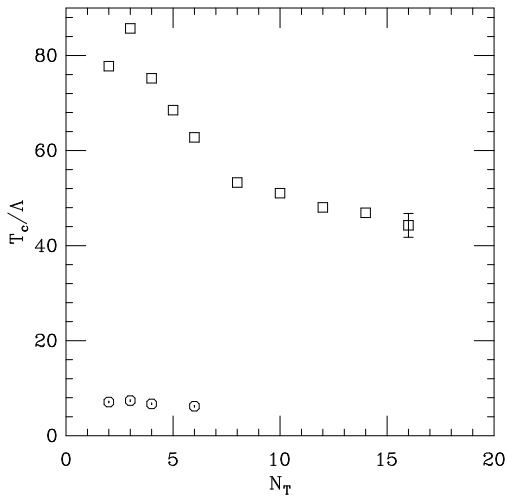

Figure 4. $T_c/\Lambda$ for the Wilson (squares) and candidate AFP (octagons) actions.

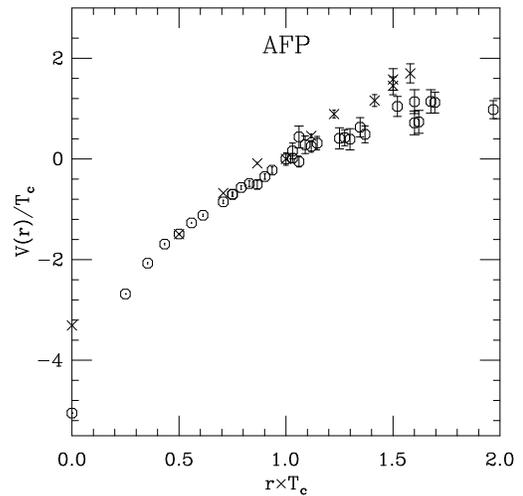

Figure 5. Potential $V(r)/T$ vs. $rT$ for the AFP action at $\beta_c(N_T = 2)$ (crosses) and $\beta_c(N_T = 4)$ (octagons).

## 5. SUMMARY

This project is incomplete. At present we have one candidate AFP action which costs a factor of 9 in computer time compared to the Wilson action. It shows asymptotic scaling in its bare coupling with a small $T_c/\Lambda$ ratio and its potential (measured with naive operators) shows better rotational invariance than the Wilson action. These are however not scaling tests. For the potential the AFP action shows qualitatively better scaling. The only quantitative scaling test we have completed so far shows excellent scaling both for the AFP and Wilson actions as well. That shows the importance of testing the scaling behaviour of several different operators. We need other tests of scaling and we need to develop the formalism of improved operators to test the potential. We are testing another RG transformation and its AFP action. Our long term goal is to apply our techniques to full QCD.

This work was supported by the U.S. Department of Energy and by the National Science Foundation and by the Swiss National Science Foundation.